\documentclass[aps,showpacs,preprintnumbers,12pt,floatfix,tightenlines]{revtex4-1}

\usepackage{latexsym} 
\usepackage{amssymb}  
\usepackage{amsfonts} 
\usepackage{amsbsy}   
\usepackage{amsmath} 
\usepackage{multirow}  
\usepackage{bm}   

\newcommand{\<}{\langle}
\renewcommand{\>}{\rangle} 

\newcommand\eqn[1]{(\ref{#1})}      
\newcommand\Eqn[1]{Eq.~(\ref{#1})}  
\newcommand{\e}{{\rm e}}   
\newcommand{\ri}{{\rm i}}

\newcommand{\nn}{\nonumber}

\newcommand{\cC}{\ensuremath{\mathcal{C}}}
\newcommand{\cP}{\ensuremath{\mathcal{P}}}
\newcommand{\cT}{\ensuremath{\mathcal{T}}}

\makeatletter 

\def\appendix{\par                              
    \setcounter{section}{0}                     
    \setcounter{subsection}{0}
    \renewcommand{\theequation}{\Alph{section}.\arabic{equation}}
    \renewcommand{\thesection}{Appendix \Alph{section}
                \setcounter{equation}{0}  } 
    \renewcommand{\thesubsection}{\Alph{section}.\arabic{subsection}}
}

\def\applabel#1{\@bsphack
  \protected@write\@auxout{}%
         {\string\newlabel{#1}{{\Alph{section}}{\thepage}}}%
  \@esphack}

\def\section{
\setcounter{equation}{0}        
\@startsection {section}{1}{\z@}{-3.5ex plus -1ex minus
 -.2ex}{2.3ex plus .2ex}{\large\bf}}
\renewcommand{\theequation}{\arabic{section}.\arabic{equation}}

\def\subsection{\@startsection{subsection}{2}{\z@}{-3.25ex plus -1ex minus
 -.2ex}{1.5ex plus .2ex}{\normalsize\bf}}

\def\subsubsection{\@startsection{subsubsection}{3}{\z@}{-3.25ex plus
 -1ex minus -.2ex}{1.5ex plus .2ex}{\normalsize}}

\makeatother   

\begin{document}

\title{\bf \cP\cT\ Symmetry as a Generalization of Hermiticity}

\author{Qing-hai Wang}
\affiliation{Department of Physics, National University of Singapore,
Singapore, 117542}
\author{Song-zhi Chia}
\author{Jie-hong Zhang}
\affiliation{NUS High School of Mathematics and Science, Singapore, 129957}

\date{April 16, 2010}

\begin{abstract}
The Hilbert space in \cP\cT-symmetric quantum mechanics is formulated as a linear vector space with a dynamic inner product. The most general \cP\cT-symmetric matrix Hamiltonians are constructed for $2\times 2$ and $3\times 3$ cases. In the former case, the \cP\cT-symmetric Hamiltonian represents the most general matrix Hamiltonian with a real spectrum. In both cases, Hermitian matrices are shown to be special cases of \cP\cT-symmetric matrices. This finding confirms and strengthens the early belief that the \cP\cT-symmetric quantum mechanics is a generalization of the conventional Hermitian quantum mechanics.
\end{abstract}

\pacs{03.65.-w, 11.30.Er}
\begin{titlepage}
\maketitle
\renewcommand{\thepage}{}          
\end{titlepage}

\section{Introduction}

The seminal paper by Bender and Boettcher in 1998 has lead to an alternative formulation of quantum mechanics (QM): non-Hermitian \cP\cT-symmetric QM \cite{BB}. Although the Hamiltonians involved appear to be not Hermitian ($H\neq H^\dag$), they yield only real spectra and the time evolution is unitary when the \cP\cT\ symmetry is not broken \cite{BB,BBJ,Review,AliReview}. 

Since the introduction of \cP\cT-symmetric QM, there has been a debate about whether \cP\cT-symmetric QM is more general than the conventional QM. In the first paper about \cP\cT\ symmetry, Bender and Boettcher stressed that \cP\cT\ symmetry is a weaker condition than Hermiticity and a \cP\cT-symmetric theory can be considered as a complex extension of a conventional Hermitian theory \cite{BB}. In 2002, Bender {\it et al.} constructed a positive-definite norm of \cP\cT-symmetric QM using the newly discovered \cC\ operator \cite{BBJ}. The claim that \cP\cT\ symmetry is more general than Hermiticity was restated in the paper's title: {\it ``Complex Extension of Quantum Mechanics.''} Interestingly, in the same year, Mostafazadeh pointed out that all pseudo-Hermitian Hamiltonians are Hermitian (self-adjoint) with respect to a positive-semidefinite inner product \cite{AliJMP2002}. As such a \cP\cT-symmetric Hamiltonian can be considered as a special case of a pseudo-Hermitian Hamiltonian with a real spectrum. In 2003, Mostafazadeh showed that diagonalizable pseudo-Hermitian Hamiltonians are extensions of \cP\cT- or \cC\cP\cT-symmetric Hamiltonians \cite{AliJMP2003}. A less general result was obtained independently by Bender {\it et al.} in Ref.~\cite{BMWparity}. But they took a different point of view which can be seen from the title of the paper: {\it ``All Hermitian Hamiltonians Have Parity.''}

The study of finite-dimensional matrix Hamiltonians may shed some light on the debate. The original $2\times 2$ \cP\cT-symmetric matrix Hamiltonian introduced in Ref.~\cite{BBJ} has three real parameters. In 2003, Bender {\it et al.} extended it to a four-parameter class and to higher dimensions \cite{BMW}. From their results, they identified \cP\cT-symmetric Hamiltonians and Hermitian Hamiltonians as two distinct extensions of real-symmetric Hamiltonians. For matrix dimension higher than two, \cP\cT-symmetric Hamiltonians have less real parameters than Hermitian Hamiltonians in the same dimension. This view was challenged immediately by Mostafazadeh in a paper titled {\it ``Exact \cP\cT-Symmetry Is Equivalent to Hermiticity''} \cite{AliJPA2003}. In this paper, the author constructed a $2\times 2$ \cP\cT-symmetric Hamiltonian with five real parameters. In 2007, Mostafazadeh and \"Oz\c{c}elik explicitly constructed the most general $2\times 2$ quasi-Hermitian Hamiltonian with six real parameters \cite{AliTurk2007}.

In this paper, we are trying to settle the debate by constructing the most general \cP\cT-symmetric matrix Hamiltonians. Part of the reason for the debate is due to the different usage of the terminology. To avoid further confusion, we restrict the term ``Hermitian conjugate'' or ``$\dag$'' in the Dirac sense: complex conjugate and transpose. In the present paper, we formulate \cP\cT-symmetric QM slightly different from the literature in two aspects. First, we define the time reversal operator as the Dirac conjugation rather than just the complex conjugation used in Ref.~\cite{BB,BBJ,BMW}. Second, we do not introduce any biorthonormal basis as in Ref.~\cite{AliJMP2002,AliJPA2003}. We link the \cC\cP\cT-inner product to the general inner product in a linear vector space with a weight function. The latter is the standard notation which can be found in modern quantum mechanics textbooks such as Ref.~\cite{BalBook}.

With this formulation, we solve for the general \cP\ operator and present the solutions explicitly in the case of $2\times 2$ and $3\times 3$. Using the general \cP\ operators, we construct the general \cP\cT-symmetric matrix Hamiltonians. We confirm that the most general $2\times 2$ \cP\cT-symmetric Hamiltonian has six real parameters as shown in Ref.~\cite{AliTurk2007}. We find that the general $3\times 3$ \cP\cT-symmetric Hamiltonian has thirteen real parameters. In the case of $2\times 2$, the general \cP\cT-symmetric Hamiltonian represents the most general matrix Hamiltonian with a real spectrum. Interestingly, this is not true in the case of $3\times 3$. In both cases, we show clearly that all Hermitian Hamiltonians are just special cases of \cP\cT-symmetric Hamiltonians. From these finite dimensional results, we conjecture that \cP\cT\ symmetry is a generalization of Hermiticity in general. 

The paper is organized as follows. In Sec.~\ref{sec:PTQM}, we give our formulation of \cP\cT-symmetric QM. In Sec.~\ref{sec:2x2}, we illustrate our ideas in the case of $2\times 2$. All relevant operators or matrices are calculated explicitly. Special cases discussed in the literature are analyzed. In Sec.~\ref{sec:3x3}, we construct the \cP\ operator and the \cP\cT-symmetric Hamiltonian in the case of $3\times 3$. Finally, in Sec.~\ref{sec:conclusion}, we give some concluding remarks.

\section{\cP\cT-Symmetric Quantum Mechanics}
\label{sec:PTQM}

In this section, we formulate \cP\cT-symmetric QM. We start with a brief summary on the inner product in QM. We adopt the notation used in Ref.~\cite{BalBook}.

\subsection{Inner Product}
\label{sec:inner}

In quantum mechanics, the Hilbert space can be considered as a linear vector space associated with an inner product. The inner product between two quantum states, denoted $(\cdot,\cdot)$, must satisfy the following conditions \cite{BalBook}:
\begin{enumerate}
\item $(\psi,\phi)$ is a complex number,
\item $(\psi,\phi)=(\phi,\psi)^*$, where $*$ denotes complex conjugate,
\item $(\psi,c_1\phi_1+c_2\phi_2)=c_1(\psi,\phi_1)+c_2(\psi,\phi_2)$, where $c_1$ and $c_2$ are complex numbers,
\item $(\phi,\phi)\geq 0$, with equality holding if and only if $\phi=0$.
\end{enumerate}
In general, we may define the inner product as
\begin{equation}
(\psi,\phi)\equiv \<\psi|W|\phi\>,
\label{eqn:inner}
\end{equation}
where $W$ is the weight function and the bra state is defined as the Hermitian conjugate of the ket state, $\<\cdot|\equiv |\cdot\>^\dag$ \cite{BalBook}.

From \Eqn{eqn:inner} and the first three properties of the inner product, it can be easily shown that $W$ must be a Hermitian operator: $W=W^\dag$. From the fourth property of the inner product, $W$ has to be positive definite. That is, all the eigenvalues of $W$ are positive.

In the Hilbert space defined above, a self-adjoint operator, such as the Hamiltonian $H$, satisfies
\begin{equation}
(\psi,H\phi)=(H\psi,\phi)
\label{eqn:self-a}
\end{equation}
for arbitrary states $\phi$ and $\psi$. All eigenvalues of a self-adjoint operator are real. And the eigenstates corresponding to different eigenvalues are orthogonal \cite{BalBook}.

Plugging the self-adjoint condition \eqn{eqn:self-a} into the definition of the inner product in \eqn{eqn:inner}, we obtain
\begin{equation}
WH=H^\dag W.
\label{eqn:weight}
\end{equation}
We may consider this equation as the definition of the weight function $W$ for a given Hamiltonian $H$. Thus, the inner product is dynamic (Hamiltonian dependent) in general.

In conventional QM, the weight function can be chosen as the identity operator. In this case, the self-adjoint condition in \eqn{eqn:weight} reduces to the Hermiticity condition: $H=H^\dag$. Since the identity operator is independent on Hamiltonians, the inner product is no longer dynamic.

\subsection{\cP\cT~Symmetry}
\label{sec:PT}

We define the time reversal operator, \cT\ as Dirac conjugate. That is, for an operator $A$,
\begin{equation}
\cT A \cT = A^\dag.
\end{equation}
It follows that $\cT^2=\openone$, here $\openone$ is the identity matrix. Note that our definition of the \cT\ operator differs from Refs.\cite{BB,BBJ,BMW}. This definition allows us to have a more general parity operator.

For the parity operator \cP, we demand it to commute with the time reversal operator and to be an involution. That is,
\begin{enumerate}
\item $[\cP,\cT]=0$, or equivalently, $\cP=\cP^\dag$,
\item $\cP^2=\openone$.
\end{enumerate}
There are obviously two trivial solutions to these constraints,
\begin{equation}
\cP_0=\pm\openone.
\label{eqn:P0}
\end{equation}
We will discuss the non-trivial $2\times 2$ solutions of \cP\ in Sec.~\ref{sec:2x2} and $3\times 3$ in Sec.~\ref{sec:3x3}.

A \cP\cT-symmetric Hamiltonian commutes with the combination operator \cP\cT:
\begin{equation}
[H,\cP\cT]=0 \qquad \Leftrightarrow \qquad \cP H^\dag \cP = H.
\label{eqn:PT-H}
\end{equation}
Using the trivial solutions of $\cP_0$ in \eqn{eqn:P0}, we get that a ${\cal P}_0{\cal T}$-symmetric Hamiltonian is Hermitian
\begin{equation}
[H,{\cal P}_0{\cal T}]=0 \qquad \Leftrightarrow \qquad H=H^\dag.
\end{equation}
In this sense, a Hermitian Hamiltonian is $\cP_0\cT$-symmetric with $\cP_0$ to be the plus or minus identity matrix.

In general, the eigenstates of a \cP\cT-symmetric Hamiltonian with different eigenvalues are {\it not}\ orthogonal with respect to the Dirac inner product. To solve this problem, one may define a \cP\cT\ inner product as
\begin{equation}
(\psi,\phi)_{\cP\cT} \equiv \<\psi|\cP|\phi\>.
\end{equation}
From the commutation relation between $H$ and \cP\cT, it can be shown that the eigenstates of $H$ with different eigenvalues are orthogonal with respect to the \cP\cT\ inner product \cite{BB,Review}.

However, the norm with respect to the \cP\cT\ inner product is not positive definite. This is simply because \cP\ has negative eigenvalues. One has to normalize the eigenstates to $\pm 1$. To overcome this difficulty, one needs to find a positive-definite norm by introducing the \cC\ operator \cite{BBJ}. Here, we define the \cC\ operator as
\begin{equation}
\cC \equiv \sum_i |E_i\>\<E_i|\cP,
\label{eqn:C}
\end{equation}
where $|E_i\>$ are the eigenstates of $H$ with the \cP\cT-norm $+1$ or $-1$.

Our definition of the \cC\ operator has the same properties as the one constructed in Ref.~\cite{BBJ}. From the orthogonality and the (non-positive-definite) normalization of the \cP\cT\ inner product, it can be shown that $|E_i\>$ are eigenstates of \cC\ with eigenvalue equal to the \cP\cT-norm:
\begin{equation}
\cC|E_i\>=\<E_i|\cP|E_i\>|E_i\>.
\label{eqn:CE}
\end{equation}
Thus, \cC\ commutes with $H$. The \cC\ operator also commutes with \cP\cT: $[\cC,\cP\cT]=0$. This fact can be verified by using $\cP\cC^\dag\cP=\cC$. Because the eigenvalues of $\cC^2$ are all unity, the \cC\ operator is an involution just like the \cP\ operator.

Note that we do not require $|E_i\>$ to be simultaneous eigenstates of \cP\cT\ and of $H$. In fact, this can only be achieved in the special cases of symmetric matrices, such as Hamiltonians constructed in Refs.~\cite{BBJ,BMW}. We will discuss more details about these two examples in Sec.\ref{sec:2x2}.

Equipped with the \cC\ operator, we are ready to construct an inner product with the positive-definite norm. We define the \cC\cP\cT\ inner product as
\begin{equation}
(\psi,\phi)_{\cC\cP\cT} \equiv \<\psi|\cP\cC|\phi\>.
\end{equation}
Comparing to the general inner product in \eqn{eqn:inner}, we recognize the weight function for the \cC\cP\cT-inner product is
\begin{equation}
W=\cP\cC.
\end{equation}

Since the weight function is positive-definite, we may find its square root $W=\eta^2$, where $\eta$ is Hermitian, $\eta^\dag=\eta$. Using the operator $\eta$, we may define a Hermitian Hamiltonian $h$, which has same spectrum as a \cP\cT-symmetric Hamiltonian $H$ \cite{AliJPA2003},
\begin{equation}
h \equiv \eta H\eta^{-1}, \quad {\rm where}\quad h=h^\dag.
\label{eqn:h}
\end{equation}
We would like to emphasize that the transformation $\eta$ is Hermitian rather than unitary, hence the above relation is not unitary equivalence in the usual sense.

\section{$2\times 2$ Case}
\label{sec:2x2}

In this section, we illustrate the ideas in the previous section by using $2\times 2$ matrices. We find that any $2\times 2$ Hermitian matrix is a special case of the general \cP\cT-symmetric matrices. Furthermore, it is found that the most general matrix with a real spectrum must coincide with the general \cP\cT-symmetric matrix we constructed.

Since the \cP\ operator is an involution, it is a square root of the identity matrix. In $2\times 2$ matrices, other than the trivial roots in \eqn{eqn:P0}, there is a non-trivial root with the form
\begin{equation}
\cP=\left(
\begin{array}{cc}
\cos\theta & \sin\theta~ \e^{-\ri\varphi}\\
\sin\theta~ \e^{\ri\varphi} & -\cos\theta
\end{array}
\right),
\label{eqn:P_2x2}
\end{equation}
where $\theta$ and $\varphi$ are two real parameters. In terms of the Pauli matrices, the \cP\ operator can be written as $\cP={\bf n}^r\cdot\bm{\sigma}$, where ${\bf n}^r\equiv (\sin\theta\cos\varphi,\sin\theta\sin\varphi,\cos\theta)$
is a unit vector.

To find the general \cP\cT-symmetric Hamiltonian, we use the following {\it ansatz}:
\begin{equation}
H=\varepsilon\openone + \bm{\alpha}\cdot\bm{\sigma},
\label{eqn:H_ansatz}
\end{equation}
where $\varepsilon$ and $\bm{\alpha}\equiv(\alpha_x,\alpha_y,\alpha_z)$ are complex numbers. Plugging the above {\it ansatz}\ into \eqn{eqn:PT-H}, we get the equations satisfied by $\varepsilon$ and $\bm{\alpha}$:
\begin{equation}
\varepsilon = \varepsilon^*, \qquad
\bm{\alpha}+\bm{\alpha}^* = 2 \left(\bm{\alpha}^*\cdot{\bf n}^r\right) {\bf n}^r.
\label{eqn:alpha}
\end{equation}
The first equation simply says that $\varepsilon$ is real. The second equation can be written as
\begin{equation}
\sum_i M_{ki}\alpha_i = \alpha_k^*,\quad {\rm with}\quad M_{ki} \equiv -\delta_{ki} +2 n^r_k n^r_i.
\end{equation}
In this form, the searching for a \cP\cT-symmetric Hamiltonian becomes an eigenvalue problem. If we separate the real part and the imaginary part as $\bm{\alpha}={\bf A}+\ri\,{\bf B}$, then ${\bf A}$ and ${\bf B}$ can be considered as the eigenvectors of the matrix $M$ with eigenvalues $+1$ and $-1$, respectively. The matrix $M$ always has one eigenvector with eigenvalue $+1$, and two eigenvectors with eigenvalue $-1$. It is easy to show that the eigenvector with positive eigenvalue is parallel to ${\bf n}^r$ and the eigenvectors with negative eigenvalue are perpendicular to ${\bf n}^r$. Thus, the solutions to the eigenvalue problem are
\begin{equation}
{\bf A} = \gamma {\bf n}^r,\qquad
{\bf B} = \mu {\bf n}^\theta +\nu {\bf n}^\varphi,
\label{eqn:AB}
\end{equation}
where $\gamma$, $\mu$, and $\nu$ are real parameters and ${\bf n}^\theta \equiv (\cos\theta\cos\varphi,\cos\theta\sin\varphi,-\sin\theta)$ and ${\bf n}^\varphi \equiv (-\sin\varphi,\cos\varphi,0)$ are two unit vectors.

Thereafter, plugging $\bm{\alpha} = \gamma {\bf n}^r + \ri\mu {\bf n}^\theta + \ri\nu {\bf n}^\varphi$ into the {\it ansatz}\ in \eqn{eqn:H_ansatz}, we get the general \cP\cT-symmetric $2\times 2$ Hamiltonian matrix,
\begin{eqnarray}
H&=&\varepsilon\openone + \left(\gamma {\bf n}^r + \ri\mu {\bf n}^\theta + \ri\nu {\bf n}^\varphi\right)\cdot\bm{\sigma}\nn\\
&=&\left(
\begin{array}{cc}
\varepsilon + \gamma\cos\theta - \ri\mu\sin\theta &
(\gamma\sin\theta + \ri\mu\cos\theta+\nu) \e^{-\ri\varphi}\\
(\gamma\sin\theta + \ri\mu\cos\theta-\nu) \e^{\ri\varphi} &
\varepsilon - \gamma\cos\theta + \ri\mu\sin\theta
\end{array}
\right).
\label{eqn:H}
\end{eqnarray}
This Hamiltonian has six real parameters: $\varepsilon$, $\gamma$, $\mu$, $\nu$, $\theta$, and $\varphi$. All $2\times 2$ Hermitian matrices can be recovered as special cases with $\mu=\nu=0$. In other words, the \cP\cT\ symmetry is a generalization of Hermiticity.

The eigenvalues of the Hamiltonian $H$ are
\begin{equation}
E_\pm=\varepsilon \pm \sqrt{\gamma^2-\mu^2-\nu^2}.
\label{eqn:Epm}
\end{equation}
For eigenvalues to be real, it requires $\gamma^2 \geq \mu^2 + \nu^2$. For simplicity, we only consider the non-degenerated case with $\gamma^2 > \mu^2 + \nu^2$ in this paper. The corresponding eigenstates are
\begin{equation}
|E_\pm\> = \frac{u}{\sqrt{2}} \left(
\begin{array}{c}
\e^{\ri\kappa_0-\ri\varphi} \sqrt{1 + \frac{\nu}{\gamma}\sin\theta \pm \frac{1}{\gamma}\sqrt{\gamma^2-\mu^2-\nu^2} \cos\theta}\\
\e^{\ri\kappa_\pm} \sqrt{1 - \frac{\nu}{\gamma}\sin\theta \mp \frac{1}{\gamma}\sqrt{\gamma^2-\mu^2-\nu^2} \cos\theta}
\end{array}\right),
\label{eqn:eigenstates}
\end{equation}
where we have defined three angles and a normalization constant as
\begin{eqnarray}
\kappa_0 &\equiv& \arg \left(\gamma\sin\theta  + \nu + \ri \mu \cos\theta \right),\nn\\
\kappa_\pm &\equiv& \arg \left(-\gamma\cos\theta \pm \sqrt{\gamma^2-\mu^2-\nu^2}  + \ri \mu \sin\theta \right),\nn\\
u &\equiv& \sqrt{\frac{\gamma^2}{\gamma^2-\mu^2-\nu^2}}.
\end{eqnarray}
The normalization is chosen such that
\begin{equation}
\<E_\pm|\cP|E_\pm\>=\pm {\rm sign}(\gamma).
\end{equation}

The six-parameter-class of matrix in \eqn{eqn:H} with the condition $\gamma^2 \geq \mu^2 + \nu^2$ coincides with the most general $2\times 2$ matrix with only real eigenvalues. Qualitatively, this can be seen from parameter counting. A general complex $2\times 2$ matrix has eight real parameters. The reallity of all eigenvalues puts two constraints on the matrix. Therefore, the most general $2\times 2$ matrix with only real eigenvalues should consist of six real parameters.

This coincidence can also be proved rigorously. By direct computation, the eigenvalues of an arbitrary $2\times 2$ matrix with the form of the {\it ansatz}\ in \eqn{eqn:H_ansatz} are $E_\pm=\varepsilon \pm \sqrt{\bm{\alpha}\cdot\bm{\alpha}}$.
Imposing reality condition on the eigenvalues leads to
\begin{equation}
\varepsilon = \varepsilon^*, \qquad \bm{\alpha}\cdot\bm{\alpha} \geq 0.
\label{eqn:MostGeneral}
\end{equation}
The first condition is the same constraint on the parameter $\varepsilon$ as in the \cP\cT-symmetric Hamiltonian in \eqn{eqn:H}. The second condition in \eqn{eqn:MostGeneral} implies that the real part and the imaginary part of $\bm{\alpha}$ are perpendicular to each other, ${\bf A}\cdot{\bf B}=0$, and that the real part vector is not shorter than the imaginary part vector, ${\bf A}\cdot{\bf A}\geq {\bf B}\cdot{\bf B}$. Without loss of generality, we may parametrize the real part vector as ${\bf A} = \gamma {\bf n}^r$. Then the above conditions lead to a unique solution for the imaginary part vector which can be parametrized as ${\bf B} = \mu {\bf n}^\theta +\nu {\bf n}^\varphi$. Since ${\bf A}$ is not shorter than ${\bf B}$, we have $\gamma^2 \geq \mu^2 + \nu^2$. Clearly then, the \cP\cT-symmetric Hamiltonian in \eqn{eqn:H} represents the most general $2\times 2$ matrices with only real eigenvalues.

For the \cP\cT-symmetric Hamiltonian $H$ in \eqn{eqn:H}, the \cC\ operator can be calculated directly from its definition in \eqn{eqn:C}. It is also straightforward to construct it from the relation in \eqn{eqn:CE}. Either way, the \cC\ operator is found to be
\begin{equation}
\cC = \frac{u}{\gamma}\bm{\alpha}\cdot \bm{\sigma}.
\end{equation}
The form of the \cC\ operator is not a surprise because it is defined as an involution. Therefore, it must be a square root of the identity matrix just like the \cP\ operator. This fact can be easily verified by observing that $(u/\gamma)\bm{\alpha}$ is a unit vector.

Because the \cC\ operator has eigenvalues $\cC|E_\pm\> = \pm {\rm sign} (\gamma)|E_\pm\>$, the eigenstates in \eqn{eqn:eigenstates} are normalized to unity with respect to the \cC\cP\cT-inner product. Thus, we have a set of orthonormal eigenstates,
\begin{equation}
\<E_i|\cP\cC|E_j\>=\delta_{ij}, \qquad i,j=\pm.
\end{equation}

The weight function has the form
\begin{equation}
W = \cP\cC = u\left(\openone + \bm{\beta}\cdot\bm{\sigma} \right), \qquad {\rm where}\quad \bm{\beta} \equiv \frac{\nu}{\gamma} {\bf n}^\theta - \frac{\mu}{\gamma} {\bf n}^\varphi.
\end{equation}
Note that $\bm{\beta}$ is a unit vector and it is perpendicular to $\bm{\alpha}$. Interestingly, the square root of $W$ has the form
\begin{equation}
\eta_\pm=\frac{1}{\sqrt{2(u\pm 1)}}(W\pm\openone).
\label{eqn:eta}
\end{equation}
There are two solutions for $\eta$, which is {\it not}\ because of the arbitrary overall sign in the square root. Rather, it is corresponding to two choices of mapping the eigenstates during the similarity transformation in \eqn{eqn:h}. In the Hermitian limit, $W\to\openone$, $\eta$ is once again a square root of the identity matrix. The ``$+$'' sign in \eqn{eqn:eta} has the limit $\eta_+\to\openone$ and the ``$-$'' sign has the limit $\eta_-\to \bm{\beta}\cdot\bm{\sigma}.$

Using the operator $\eta$, we find the Hermitian equivalence of \cP\cT-symmetric Hamiltonian $H$ in \eqn{eqn:H},
\begin{equation}
h = \varepsilon \openone \pm \frac{\gamma}{u} {\bf n}^r\cdot \bm{\sigma}.
\end{equation}

Now let us consider some special cases. We show that both Hermitian Hamiltonians and several \cP\cT-symmetric $2\times 2$ Hamiltonians studied in the literature can be reduced from our general \cP\cT-symmetric Hamiltonian in \eqn{eqn:H}.

\subsection{Special Case 1: Hermiticity}
\label{sec:Hermitian}

If we set $\mu=\nu=0$, $H$ becomes Hermitian, $H=H^\dag$,
\begin{equation}
H_{\rm Hermitian}=\varepsilon\openone + \gamma {\bf n}^r\cdot\bm{\sigma}=\left(
\begin{array}{cc}
\varepsilon + \gamma\cos\theta & \gamma\sin\theta~ \e^{-\ri\varphi}\\
\gamma\sin\theta~ \e^{\ri\varphi} & \varepsilon - \gamma\cos\theta
\end{array}
\right).
\end{equation}
This matrix Hamiltonian has four real parameters, and it includes all $2\times 2$ Hermitian matrices.

In this case, the weight function reduces to the identity matrix and the \cC\ operator coincides with the parity operator:
\begin{equation}
W_{\rm Hermitian}=\openone, \qquad \cC_{\rm Hermitian}=\cP.
\end{equation}

All these observations are consistent with the conventional QM. We may say that the Hermitian Hamiltonian is a special case of the \cP\cT-symmetric Hamiltonian with $\cC=\cP$. Or, equivalently, \cP\cT\ symmetry is a generalization of Hermiticity.

\subsection{Special Case 2: Bender-Brody-Jones Hamiltonian}
\label{sec:BBJ}

In Ref.~\cite{BBJ}, Bender {\it et al.} studied a \cP\cT-symmetric Hamiltonian with real and symmetric off-diagonal matrix elements. Their choice of the parity operator is $\cP_{\rm BBJ}=\sigma_x$. This case can be reduced from our general case by setting $\nu=\varphi=0$ and $\theta=\pi/2$. In particular, we have
\begin{equation}
H_{\rm BBJ}=\left(
\begin{array}{cc}
\varepsilon - \ri\mu & \gamma\\
\gamma & \varepsilon + \ri\mu
\end{array}
\right), \qquad
\cC_{\rm BBJ}=\sqrt{\frac{\gamma^2}{\gamma^2-\mu^2}}\left(
\begin{array}{cc}
 -\ri\frac{\mu}{\gamma} & 1\\
1 & \ri\frac{\mu}{\gamma}
\end{array}
\right).
\end{equation}
These expressions are equivalent to those in Ref.~\cite{BBJ} by mapping our parameters $\varepsilon$, $\mu$, and $\gamma$ to $r\cos\theta$, $-r\sin\theta$, and $s$ therein.

\subsection{Special case 3: Bender-Meisinger-Wang Hamiltonian}
\label{sec:BMW}

In Ref.~\cite{BMW}, Bender {\it et al.}\ generalized the Hamiltonian matrix in Ref.~\cite{BBJ} by choosing a one-parameter class of parity operator,
\begin{equation}
\cP_{\rm BMW}=\left(
\begin{array}{cc}
\cos\theta & \sin\theta\\
\sin\theta & -\cos\theta
\end{array}
\right).
\end{equation}
This can be recovered by setting $\varphi=0$ in \eqn{eqn:P_2x2}. If we consider only symmetric Hamiltonian as in Ref.~\cite{BMW}, we can further set $\nu=0$ in \eqn{eqn:H}. With this choice of parameters, we have
\begin{eqnarray}
H_{\rm BMW}&=&\left(
\begin{array}{cc}
\varepsilon + \gamma\cos\theta - \ri\mu\sin\theta& \gamma\sin\theta + \ri\mu\cos\theta\\
\gamma\sin\theta+ \ri\mu\cos\theta & \varepsilon - \gamma\cos\theta + \ri\mu\sin\theta
\end{array}
\right),\nn\\
\cC_{\rm BMW}&=&\sqrt{\frac{\gamma^2}{\gamma^2-\mu^2}}\left(
\begin{array}{cc}
\cos\theta -\ri\frac{\mu}{\gamma}\sin\theta& \sin\theta + \ri\frac{\mu}{\gamma}\cos\theta\\
\sin\theta+ \ri\frac{\mu}{\gamma}\cos\theta & - \cos\theta + \ri\frac{\mu}{\gamma}\sin\theta
\end{array}
\right).
\end{eqnarray}
Once again, these are the same formulas as in Ref.~\cite{BMW} by properly mapping the parameters.

The symmetric Hamiltonian, $H$ in \eqn{eqn:H} with $\nu=\varphi=0$, has additional properties. In this case, the eigenstates of $H$ are also the eigenstates of \cP\cT. By choosing a proper phase, the eigenvalue of \cP\cT\ can be set to unity:
\begin{equation}
\cP\cT|E_\pm\> \equiv \cP|E_\pm\>^* = |E_\pm\>.
\end{equation}

One may think that our definition of \cC\cP\cT-inner product is slightly different from the one in the literature. In Refs.~\cite{BBJ,BMW} the \cC\cP\cT-inner product was defined as
\begin{equation}
(\psi,\phi)_{\rm BBJ-BMW} = (\cC\cP\cT|\psi\>)^T|\phi\> = (\cC\cP|\psi\>^*)^T|\phi\>=\<\psi|\cP^T\cC^T|\phi\>,
\end{equation}
where $T$ denotes matrix transpose. This definition leads to a weight function $W_{\rm BBJ-BMW}=\cP^T\cC^T$. Since both \cP\ and \cC\ are symmetric in this case, the two definitions of the inner product are actually the same.

\subsection{Special case 4: Mostafazadeh Hamiltonian}
\label{sec:Ali2x2}

In Ref.~\cite{AliJPA2003}, Mostafazadeh introduced a five-parameter-class of Hamiltonians with the form
\begin{equation}
H_{\rm Mostafazadeh}=\left(
\begin{array}{cc}
r + t\cos\phi - \ri s\sin\phi& t\sin\phi + \ri(s\cos\phi-u)\\
t\sin\phi+ \ri(s\cos\phi+u) & r - t\cos\phi + \ri s\sin\phi
\end{array}
\right).
\end{equation}
This is a special case of our general Hamiltonian in \eqn{eqn:H} by the following parameter mapping:
\begin{eqnarray}
\varepsilon \to r, &\qquad& \gamma \to \sqrt{t^2+u^2},\nn\\
\mu \to \frac{s\sqrt{t^2+u^2}\sin\phi}{\sqrt{t^2\sin^2\phi+u^2}},
&\qquad& \nu \to -\frac{su\cos\phi}{\sqrt{t^2\sin^2\phi+u^2}},\nn\\
\cos\theta \to \frac{t\cos\phi}{\sqrt{t^2+u^2}}, &\qquad&
\tan\varphi \to \frac{u}{t\sin\phi}.
\end{eqnarray}

\subsection{Special case 5: Mostafazadeh-\"Oz\c{c}elik Hamiltonian}

In Ref.~\cite{AliTurk2007}, Mostafazadeh and \"Oz\c{c}elik constructed a six-parameter-class of Hamiltonians with a very elegant form
\begin{equation}
H_{\rm MO}=q\openone_{2\times 2} + E \left(
\begin{array}{cc}
\cos\Theta & \e^{-\ri\Phi} \sin\Theta\\
\e^{\ri\Phi} \sin\Theta & \cos\Theta
\end{array}
\right),
\label{eqn:MOH}
\end{equation}
where $q$ and $E$ are real and $\Theta$ and $\Phi$ are complex. In principle, this Hamiltonian is equivalent to our Hamiltonian in \eqn{eqn:H} by the following parameter mapping,
\begin{eqnarray}
\varepsilon \to q, &\qquad& \pm\sqrt{\gamma^2-\mu^2-\nu^2} \to E,\nn\\
\gamma\cos\theta-\ri\mu\sin\theta \to E\cos\Theta, &\qquad&
\frac{\gamma\sin\theta + \ri\mu\cos\theta - \nu}{\gamma\sin\theta + \ri\mu\cos\theta + \nu}\e^{2\ri\varphi} \to \e^{2\ri\Phi}.
\end{eqnarray}
However, at the degenerate point, $E=0$ in \eqn{eqn:MOH} or $\gamma^2=\mu^2+\nu^2$ in \eqn{eqn:H}, these two parametrizations are no longer equivalent. In this special point, $H_{\rm MO}$ is proportional to the identity matrix but $H$ in \eqn{eqn:H} is not.  

\section{$3\times 3$ Case}
\label{sec:3x3}

In this section, we reveal the general form of \cP\cT-symmetric $3\times 3$ matrix Hamiltonians. In this case, we use the Gell-Mann matrices, which are the generalization of the Pauli matrices in $3\times 3$. Any $3\times 3$ matrix can be written as a linear combination of the identity matrix and the eight Gell-Mann matrices. If a matrix is Hermitian, the expansion coefficients are all real.

There are two types of solutions for the \cP\ operator as well. The first type is the trivial solutions $\cP_0=\pm\openone$. The second type is the non-trivial solutions. If we expand the non-trivial solutions of \cP\ as
\begin{equation}
\cP_{3\times 3}=\pm\left(P_0 \openone + \sum_{i=1}^8 P_i \lambda_i\right),
\label{eqn:Pexp}
\end{equation}
where $\lambda_i$ are Gell-Mann matrices for $i=1,\cdots,8$. We find that
\begin{equation}
P_0=\frac{1}{3}
\end{equation}
and that coefficients $P_i$ depend on four independent parameters. We thus choose the parametrization as
\begin{eqnarray}
P_4&=&\sin 2\chi\sin\theta\cos\varphi,\nn\\
P_5&=&\sin 2\chi\sin\theta\sin\varphi,\nn\\
P_6&=&\sin 2\chi\cos\theta\cos\rho,\nn\\
P_7&=&\sin 2\chi\cos\theta\sin\rho.
\label{eqn:P4567}
\end{eqnarray}
The other four components, $P_1$, $P_2$, $P_3$, and $P_8$ depend on the sign of $\cos 2\chi$. For the case of $\cos 2\chi \geq 0$, we have
\begin{eqnarray}
P_1 &=& -\sin^2\chi\sin 2\theta \cos(\rho-\varphi),\nn\\
P_2 &=& \sin^2\chi\sin 2\theta \sin(\rho-\varphi),\nn\\
P_3 &=& \sin^2\chi\cos 2\theta,\nn\\
P_8 &=& \frac{1}{2\sqrt{3}}(1+3\cos 2\chi).
\label{eqn:P1238}
\end{eqnarray}
Plugging $P_0$ and $P_i$ into \eqn{eqn:Pexp}, we get the non-trivial four-parameter solutions of the \cP\ operator with the form
\begin{equation}
\cP_{3\times 3}=\pm\left(
\begin{array}{ccc}
\cos 2\chi \sin^2\theta + \cos^2\theta&
-\sin^2\chi \sin 2\theta~\e^{\ri(\rho-\varphi)}&
\sin 2\chi \sin\theta~\e^{-\ri\varphi}\\
-\sin^2\chi\sin 2\theta~\e^{-\ri(\rho-\varphi)}&
\cos 2\chi \cos^2\theta + \sin^2\theta&
\sin 2\chi \cos\theta~\e^{-\ri\rho}\\
\sin 2\chi\sin\theta~\e^{\ri\varphi} &
\sin 2\chi \cos\theta~\e^{\ri\rho}&
-\cos 2\chi
\end{array}
\right).
\label{eqn:P_3x3}
\end{equation}
For the case of $\cos 2\chi<0$, the correct results are obtained by replacing $\cos 2\chi$ by $-\cos 2\chi$ in \eqn{eqn:P1238} and \eqn{eqn:P_3x3}.

Just like in the $2\times 2$ case, \cP\cT-symmetric Hamiltonians can be found by solving an eigenvalue problem. If we use the {\it ansatz}\ of the Hamiltonian as
\begin{equation}
H_{3\times 3}=\varepsilon\openone + \sum_{i=1}^8 \alpha_i \lambda_i,
\label{eqn:H3x3}
\end{equation}
the \cP\cT\ symmetry leads to the conditions of $\varepsilon=\varepsilon^*$ and that $\alpha_i$ satisfy the eigenvalue equation
\begin{equation}
\sum_{i=1}^8 M_{ki}\alpha_i=\alpha_k^*,
\end{equation}
where
\begin{equation}
M_{ki} = P_0^2\delta_{ki} + 2P_0 \sum_j P_j d^{ijk}+\frac{2}{3} P_k P_i + \sum_{jml}P_j P_m\left( d^{ijl}d^{lmk} + f^{ijl}f^{lmk}\right)
\label{eqn:Mki3x3}
\end{equation}
with $d^{ijk}$ and $f^{ijk}$ being symmetric and antisymmetric structure constants of the SU(3) group.

Once again, the real part of $\alpha_i$ forms eigenvectors of $M$ with eigenvalue $+1$, and the imaginary part of $\alpha_i$ forms eigenvectors with eigenvalue $-1$. There are always four eigenvectors with eigenvalue $+1$ and four eigenvectors with eigenvalue $-1$.

It is straightforward to show that the vector $P_i$ with components defined in \eqn{eqn:P4567} and \eqn{eqn:P1238} is an eigenvector of $M$ with eigenvalue $+1$. All other eigenvectors can be constructed by the derivatives of $P_i$. The set of all four first order derivatives, $\{\partial_\chi P_i, \partial_\theta P_i, \partial_\rho P_i, \partial_\varphi P_i\}$, forms a subspace with eigenvalue $-1$. The remaining three eigenvectors with eigenvalue $+1$ can be constructed from the second derivatives. Below is a choice of orthonormal set of eigenvectors with eigenvalue $+1$,
\begin{eqnarray}
A_i^{(1)} &=& \frac{\sqrt{3}}{2}P_i \nn\\
          &=& \frac{\sqrt{3}}{2}\Biggl(
	-\sin^2\chi\sin 2\theta\cos(\rho-\varphi),
	\sin^2\chi\sin 2\theta\sin(\rho-\varphi),
	\sin^2\chi\cos 2\theta,\nn\\
	&&\quad \sin 2\chi\sin\theta\cos\varphi,
	\sin 2\chi\sin\theta\sin\varphi, \sin 2\chi\sin\theta\cos\rho,
	\sin 2\chi\sin\theta\sin\rho, \frac{1+3\cos 2\chi}{2\sqrt{3}}\Biggr), \nn\\
A_i^{(2)} &=& \frac{1}{2} \partial_\chi^2 P_i + \frac{3}{2} P_i \nn\\
	&=&\frac{1}{2} \left(-\frac{3+\cos 2\chi}{2}\sin 2\theta\cos(\rho-\varphi),
	\frac{3+\cos 2\chi}{2}\sin 2\theta\sin(\rho-\varphi),
	\frac{3+\cos 2\chi}{2}\cos 2\theta, \right.\nn\\
	&& \quad
	-\sin 2\chi\sin\theta\cos\varphi,
	-\sin 2\chi\sin\theta\sin\varphi,
	-\sin 2\chi\cos\theta\cos\rho,
	-\sin 2\chi\cos\theta\sin\rho,\nn\\
	&&\quad \left.
	\frac{\sqrt{3}}{2}\sin^2\chi\right), \nn\\
A_i^{(3)} &=& \Bigl(-\cos\chi\cos 2\theta\cos(\rho-\varphi),
	\cos\chi\cos 2\theta\sin(\rho-\varphi),
	-\cos\chi\sin 2\theta,\nn\\
 	&&\quad 
	-\sin\chi\cos\theta\cos\varphi, 	
	-\sin\chi\cos\theta\sin\varphi,
	\sin\chi\sin\theta\cos\rho,\sin\chi\sin\theta\sin\rho,0
	\Bigr),\nn\\
A_i^{(4)} &=& \Bigl(\cos\chi\sin(\rho-\varphi), \cos\chi\cos(\rho-\varphi),
	0, -\sin\chi\cos\theta\sin\varphi,
	\sin\chi\cos\theta\cos\varphi, \nn\\
	&&\quad 
	\sin\chi\sin\theta\sin\rho,-\sin\chi\sin\theta\cos\rho,0\Bigr).
\end{eqnarray}
Likewise, a set of eigenvectors with eigenvalue $-1$ can be chosen as
\begin{eqnarray}
B_i^{(1)} &=& \frac{1}{2} \partial_\chi P_i \nn\\
	&=& \frac{1}{2} \Bigl(-\sin 2\chi\sin 2\theta\cos(\rho-\varphi),
	\sin 2\chi\sin 2\theta\sin(\rho-\varphi),
	\sin 2\chi\cos	2\theta,2\cos 2\chi\sin\theta\cos\varphi,\nn\\
	&&\quad 2\cos 2\chi\sin\theta\sin\varphi, 2\cos 2\chi\cos\theta\cos\rho,
	2\cos 2\chi\cos\theta\sin\rho,
	-\sqrt{3}\sin 2\chi\Bigr),\nn\\
B_i^{(2)} &=& \frac{1}{2\sin\chi} \partial_\theta P_i \nn\\
	&=& \Bigl(-\sin\chi\cos 2\theta\cos(\rho-\varphi),
	\sin\chi\cos 2\theta\sin(\rho-\varphi),
	-\sin\chi\sin 2\theta,\nn\\
	&&\quad \cos\chi\cos\theta\cos\varphi,
	\cos\chi\cos\theta\sin\varphi, -\cos\chi\sin\theta\cos\rho,
	-\cos\chi\sin\theta\sin\rho,0\Bigr),\nn\\
B_i^{(3)} &=& \frac{\partial_\phi P_i+\partial_\varphi P_i}{\sin 2\chi} \nn\\
	&=& \Bigl(0,0,0,-\sin\theta\sin\varphi, \sin\theta\cos\varphi,
	-\cos\theta\sin\rho,\cos\theta\cos\rho,0\Bigr), \nn\\
B_i^{(4)} &=& \Bigl(\sin\chi\sin(\rho-\varphi),
	\sin\chi\cos(\rho-\varphi),0,
	\cos\chi\cos\theta\sin\varphi,\nn\\
	&&\quad -\cos\chi\cos\theta\cos\varphi,
	-\cos\chi\sin\theta\sin\rho, \cos\chi\sin\theta\cos\rho,0\Bigr).
\end{eqnarray}

With the solution of $\alpha_i=A_i+\ri\,B_i$, we can construct the general $3\times 3$ \cP\cT-symmetric Hamiltonian by plugging the above results into the {\it ansatz}\ in \eqn{eqn:H3x3}:
\begin{equation}
H_{3\times 3}=\varepsilon \openone + \sum_{i=1}^8 \left[\gamma_1 A_i^{(1)} +
\gamma_2 A_i^{(2)} + \gamma_3 A_i^{(3)} + \gamma_4 A_i^{(4)} + \ri\left(\mu_1 B_i^{(1)} + \mu_2 B_i^{(2)} + \mu_3 B_i^{(3)} + \mu_4 B_i^{(4)}\right)\right]\lambda_i.
\label{eqn:HH3x3}
\end{equation}
This construction has thirteen real parameters, four in \cP, four $\gamma$'s, four $\mu$'s, and one $\varepsilon$. Any $3\times 3$ Hermitian Hamiltonian can be considered as a special case with all $\mu$'s vanishing, which also has the correct number of parameters: nine. Unlike the $2\times 2$ case, the general \cP\cT-symmetric matrix Hamiltonian in \eqn{eqn:HH3x3} does not present all $3\times 3$ matrices with all real eigenvalues. This can be seen by a simple parameter counting. A $3\times 3$ matrix with all real eigenvalues should have fifteen real parameters but $H_{3\times 3}$ in \eqn{eqn:HH3x3} only has thirteen parameters.

\section{Conclusion}
\label{sec:conclusion}

In this paper, we find the general \cP\ operator and construct the general \cP\cT-symmetric matrix Hamiltonians in $2\times 2$ and $3\times 3$. In both cases, \cP\cT\ symmetry can be considered as a generalization of Hermiticity. We conjecture that this statement is true in general.

We convert the searching for a general \cP\cT-symmetric Hamiltonian problem to an eigenvalue problem. This method also applies to higher dimensions. For example, the definition for the matrix $M_{ki}$ in \eqn{eqn:Mki3x3} can be easily generalized to $N$ dimensions by replacing $3$ by $N$.

\vskip 2cm
{\bf Acknowledgments}

QW is very grateful to Professor Carl M.~Bender for helpful comments and correspondence. QW would also like to thank Dr.~Jiangbin Gong for many useful discussions and some help on writing.


\begin{thebibliography}{99}

\bibitem{BB}
C.M.~Bender and S.~Boettcher,
``{\it Real Spectra in Non-Hermitian Hamiltonians Having \cP\cT\ Symmetry},''
Phys.~Rev.~Lett.~{\bf 80}, 5243-5246 (1998).

\bibitem{BBJ}
C.M.~Bender, D.C.~Brody, and H.F.~Jones,
``{\it Complex Extension of Quantum Mechanics},''
Phys.~Rev.~Lett.~{\bf 89}, 270401 (2002) [Erratum: {\it ibid.}~{\bf 92} 119902 (2004)].

\bibitem{Review}
C.M.~Bender,
``{\it Making Sense of Non-Hermitian Hamiltonians},''
Rep.~Prog.~Phys.~{\bf 70}, 947-1018 (2007) [arXiv:hep-th/0703096].

\bibitem{AliReview}
A.~Mostafazadeh,
``{\it Pseudo-Hermitian Quantum Mechanics},''
arXiv:0810.5643.

\bibitem{AliJMP2002}
A.~Mostafazadeh,
``{\it Pseudo-Hermiticity for a Class of Nondiagonalizable Hamiltonians,}''
J.~Math.~Phys.~{\bf 43}, 6343-6352 (2002) [Erratum: {\it ibid.}~{\bf 44} 943 (2003)].

\bibitem{AliJMP2003}
A.~Mostafazadeh,
``{\it Pseudo-Hermiticity and Generalized \cP\cT- and \cC\cP\cT-Symmetries,}''
J.~Math.~Phys.~{\bf 44}, 974-989 (2003).

\bibitem{BMWparity}
C.M.~Bender, P.N.~Meisinger, and Q.~Wang,
``{\it All Hermitian Hamiltonians Have Parity},''
J.~Phys.~A: Math.~Gen.~{\bf 36}, 1029-1031 (2003) [arXiv:quant-ph/0211123].

\bibitem{BMW}
C.M.~Bender, P.N.~Meisinger, and Q.~Wang,
``{\it Finite-Dimensional \cP\cT-Symmetric Hamiltonians},''
J.~Phys.~A: Math.~Gen.~{\bf 36}, 6791-6797 (2003) [arXiv:quant-ph/0303174].


\bibitem{AliJPA2003}
A.~Mostafazadeh,
``{\it Exact \cP\cT-Symmetry Is Equivalent to Hermiticity,}''
J.~Phys.~A: Math.~Gen.~{\bf 36}, 7081-7092 (2003) [arXiv:quant-ph/0304080].

\bibitem{AliTurk2007}
A.~Mostafazadeh and S.~\"Oz\c{c}elik
``{\it Explicit Realization of Pseudo-Hermitian and Quasi-Hermitian Quantum Mechanics for Two-Level Systems}''
Turk.~J.~Phys.~{\bf 30}, 437-443 (2006) [arXiv:quant-ph/0607120].

\bibitem{BalBook}
L.E.~Ballentine,
{\it Quantum Mechanics: A Modern Development} (World Scientific, Singapore, 1998).

\end{thebibliography}
\end{document}